\documentclass[twocolumn,showpacs,preprintnumbers,aps,prl,amsmath,amssymb,superscriptaddress]{revtex4}

\usepackage{color}
\usepackage{graphicx}
\usepackage{dcolumn}
\usepackage{bm}
\usepackage{amssymb}
\usepackage[german, english]{babel}
\usepackage{dcolumn} 
\begin{document}
\title{Bose-Einstein Condensation of Strontium}
\author{Simon Stellmer}
 \affiliation{Institut f\"ur Quantenoptik und Quanteninformation (IQOQI),
\"Osterreichische Akademie der Wissenschaften, 6020 Innsbruck,
Austria}
\affiliation{Institut f\"ur Experimentalphysik und
Zentrum f\"ur Quantenphysik, Universit\"at Innsbruck,
6020 Innsbruck, Austria}
\author{Meng Khoon Tey}
\affiliation{Institut f\"ur Quantenoptik und Quanteninformation (IQOQI),
\"Osterreichische Akademie der Wissenschaften, 6020 Innsbruck,
Austria}
\author{Bo Huang}
\author{Rudolf Grimm}
 \affiliation{Institut f\"ur Quantenoptik und Quanteninformation (IQOQI),
\"Osterreichische Akademie der Wissenschaften, 6020 Innsbruck,
Austria}
\affiliation{Institut f\"ur Experimentalphysik und
Zentrum f\"ur Quantenphysik, Universit\"at Innsbruck,
6020 Innsbruck, Austria}
\author{Florian Schreck}
\affiliation{Institut f\"ur Quantenoptik und Quanteninformation (IQOQI),
\"Osterreichische Akademie der Wissenschaften, 6020 Innsbruck,
Austria}

\date{\today}


\begin{abstract}
We report on the attainment of Bose-Einstein condensation with ultracold strontium atoms. We use the $^{84}$Sr isotope, which has a low natural abundance but offers excellent scattering properties for evaporative cooling. Accumulation in a metastable state using a magnetic-trap, narrowline cooling, and straightforward evaporative cooling in an optical trap lead to pure condensates containing $1.5 \times 10^5$ atoms. This
puts $^{84}$Sr in a prime position for future experiments on quantum-degenerate gases involving atomic two-electron
systems.
\end{abstract}

\maketitle

Ultracold gases of strontium atoms have been attracting great attention for various reasons. A major driving force for the development of cooling and trapping techniques since the early 1990s \cite{Kurosu1990lca} has been the realization of ultraprecise optical clocks \cite{Ido2003rfs, Boyd2007slc, Lemonde2009olc}. Many other intriguing applications related to metrology \cite{Sorrentino2006lca}, novel schemes for quantum computation \cite{Daley2008qcw, Gorshkov2009aem}, and quantum simulators of unique many-body phenomena \cite{Hermele2009mio, Gorshkov2009tom} rely on the special properties of this species. Moreover, there is considerable interest in ultracold Sr$_2$ molecules \cite{Ciurylo2004pso, Koch2008pfc} and their possible applications for testing the time variation of fundamental constants \cite{Kotochigova2009pfa}. Ultracold plasmas \cite{Killian2007unp} represent another fascinating application of strontium atoms. Many of the possible experiments could greatly benefit from the availability of quantum-degenerate samples.

The two valence electrons of strontium and the resulting singlet and triplet states are at the heart of many of these applications. The two-electron nature also has very important consequences for cooling and trapping strategies towards degeneracy. Because of its singlet character the electronic ground state does not carry a magnetic moment. Therefore optical dipole traps \cite{Grimm2000odt} are the only option to implement evaporative cooling. Moreover, magnetic Feshbach resonances, frequently applied to tune the scattering properties of other atomic systems \cite{Chin2008fri}, are absent.
Research on Bose-Einstein condensation (BEC) and degenerate Fermi gases involving atomic two-electron systems was pioneered by the Kyoto group, using various isotopes of Yb \cite{Takasu2003ssb, Fukuhara2007dfg, Fukuhara2007bec, Fukuhara2009aof}. Very recently, a BEC of $^{40}$Ca was produced at the
Physikalisch-Technische Bundesanstalt in Braunschweig \cite{Kraft2009bec}.


Experiments towards quantum degeneracy in strontium have so far been focused on the three relatively abundant isotopes $^{86}$Sr (9.9\%), $^{87}$Sr (7.0\%), and $^{88}$Sr (82.6\%), the first and the last one being bosons. The necessary phase-space density for BEC or Fermi degeneracy could not be achieved in spite of considerable efforts \cite{Katori2001lco,Ferrari2006cos}. For the two bosonic isotopes
the scattering properties turned out to be unfavorable for evaporative cooling \cite{Ferrari2006cos}. The scattering length of $^{88}$Sr is close to zero, so that elastic collisions are almost absent. In contrast, the scattering length of $^{86}$Sr is very large, leading to detrimental three-body recombination losses. As a possible way out of this dilemma, the application of optical Feshbach resonances \cite{Chin2008fri} to tune the scattering length is currently under investigation \cite{Martinezdeescobar2009moa, Yeprivate}.

In this Letter, we report on the attainment of BEC in $^{84}$Sr. This isotope has a natural abundance of only 0.56\% and, apparently for this reason, has received little attention so far. We show that the low abundance does not represent a serious disadvantage for BEC experiments, as it can be overcome by an efficient loading scheme. Because of the favorable scattering length of $+123\,a_0$ (Bohr radius $a_0 \approx 53$\,pm) \cite{Florianprivate, Stein2008fts, Martinezdeescobar2008tpp} there is no need of Feshbach tuning, and we can easily produce BECs containing $1.5\times10^5$ atoms.

Our experimental procedure
can be divided into three main stages. In the first stage, the atoms are accumulated in a magnetic trap, using a continuous loading scheme based on optical pumping into a metastable state. In the second stage, the atoms are first pumped back into the electronic ground state, laser cooled using a narrow intercombination line, and loaded into an optical dipole trap (ODT).
In the third stage, evaporative cooling is performed by lowering the depth of the ODT and, thanks to the excellent starting conditions and collision properties, BEC is attained in a straightforward way.


\begin{figure}
\includegraphics[width=85 mm]{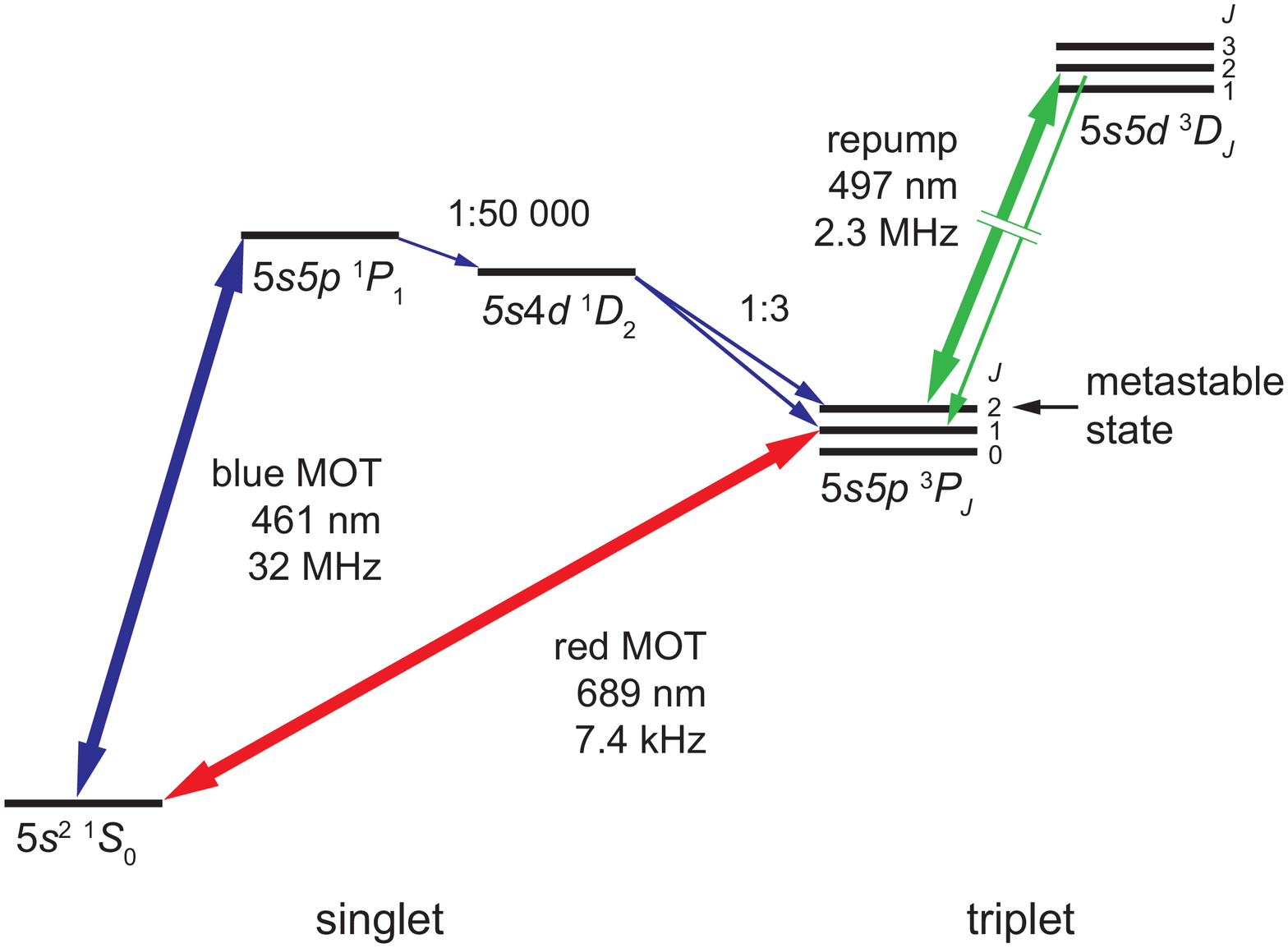}
\caption{\label{fig:Fig1LevelScheme} Schematic illustration of the energy levels and transitions used for cooling and trapping of Sr atoms. The blue MOT is operated on the strong $^1S_0$-$^1P_1$ transition. The loading of the magnetic trap proceeds via the weak leak of the excited state (branching ratio 1:50\,000) into the $^1D_2$ state, which itself decays with a 1:3 probability into the metastable $^3P_2$ state. Here the atoms can be magnetically trapped and accumulated for a long time. The transition $^3P_2$-$^3D_2$ allows us to depopulate the metastable state by transferring the atoms into the $^3P_1$ state. The latter represents the excited state of the $^1S_0$-$^3P_1$ intercombination line, used for narrowline cooling in the red MOT.}
\end{figure}

The accumulation stage takes advantage of magnetically trapped atoms in the metastable triplet state $^3P_2$; see Fig.~\ref{fig:Fig1LevelScheme}. Remarkably, such atoms are automatically produced \cite{Katori2001lco, Loftus2002mto, Xu2003cat, Nagel2003mto, Ferrari2006cos, Traverso2009iae, Mickelson2009ras} when a standard magneto-optical trap (MOT) is operated on the strong $^1S_0$-$^1P_1$ transition at a wavelength of 461\,nm \cite{MOTdetails}. A weak leak of the excited state out of the cooling cycle of this ``blue MOT'' continuously populates the metastable state and the atoms are trapped in the magnetic quadrupole field of the MOT. This continuous magnetic-trap loading mechanism is our essential tool to prepare a sufficiently large number of $^{84}$Sr atoms despite the low natural abundance of this isotope. With a steady-state number of about $3 \times 10^5$ atoms in the blue MOT, we can reach an estimated number of roughly $10^8$ atoms in the magnetic trap after typically 10\,s of loading. This enormous gain is facilitated by the long lifetime of about 35\,s for the magnetically trapped atoms under our ultrahigh vacuum conditions, which is about 3 orders of magnitude larger than the leak time constant of the blue MOT. Note that the same scheme has been applied to increase the number of $^{84}$Sr atoms for spectroscopic measurements \cite{Mickelson2009ras}. Also note that a very similar loading scheme was crucial for the attainment of BEC in Cr \cite{Griesmaier2005bec}.


In the narrowline cooling stage, a MOT is operated on the $^1S_0$-$^3P_1$ intercombination line (wavelength 689\,nm, linewidth 7.4\,kHz) using a scheme pioneered by Katori {\em et al.} \cite{Katori1999mot}, which has become an almost standard tool for the preparation of ultracold Sr. Loading of this ``red MOT'' \cite{689nmlaser} is accomplished by pumping the atoms out of the metastable reservoir using a flash of laser light resonant with the  $^3P_2$-$^3D_2$ transition at 497\,nm \cite{Sorrentino2006lca}; see Fig.~\ref{fig:Fig1LevelScheme}.
In the initial transfer phase, the magnetic field gradient is reduced from 61\,G/cm as used for the magnetic trap to 3.6\,G/cm within about 0.1\,ms. To increase the capture velocity of the red MOT we frequency modulate the light, producing sidebands that cover a detuning range between $-250$\,kHz and $-6.5$\,MHz with a spacing of 35\,kHz; here each of the MOT beams has a waist of 5\,mm and a peak intensity of 10\,mW/cm$^2$. In a compression phase, the red MOT is then slowly converted to single-frequency operation with a detuning of about $-800$\,kHz by ramping down the frequency modulation within 300\,ms.
At the same time the intensity of the MOT beams is reduced to 90\,$\mu$W/cm$^2$ and the magnetic field gradient is increased to 10.4\,G/cm. At this point, we obtain $2.5 \times 10^7$ atoms at a temperature of 2.5\,$\mu$K in an oblate cloud with diameters of 1.6\,mm horizontally and 0.4\,mm vertically.


To prepare the evaporative cooling stage, the atoms are transferred into a
crossed-beam ODT, which is derived from a 16-W laser source operating at 1030\,nm in a single longitudinal mode. Our trapping geometry follows the basic concept successfully applied in experiments on Yb and Ca BEC \cite{Takasu2003ssb, Fukuhara2007bec, Fukuhara2009aof, Kraft2009bec}. The trap consists of a horizontal and a vertical \cite{verticalnote} beam with waists of 32\,$\mu$m and 80\,$\mu$m, respectively. Initially the horizontal beam has a power of 3\,W, which corresponds to a potential depth of 110\,$\mu$K and oscillation frequencies of 1\,kHz radially and a few Hz axially. The vertical beam has 6.6\,W, which corresponds to a potential depth of 37$\,\mu$K and a radial trap frequency of 250\,Hz. Axially, the vertical beam does not provide any confinement against gravity. In the crossing region the resulting potential represents a nearly cylindrical trap \cite{nointerference}. In addition the horizontal beam provides an outer trapping region of much larger volume, which is of advantage for the trap loading.

The dipole trap is switched on at the beginning of the red MOT
compression phase. After switching off the red MOT, we observe $2.5\,\times 10^6$ atoms in the ODT with about $10^6$ of them residing in the crossed region.
At this point we measure a temperature of $\sim$10\,$\mu$K, which corresponds to roughly one tenth of the potential depth and thus points to plain evaporation in the transfer phase. We then apply forced evaporative cooling
by exponentially reducing the power of both beams with a $1/e$ time constant of $\sim$\,3\,s \cite{robustramp}. The evaporation process starts under excellent conditions, with a peak number density of $1.2 \times 10^{14}$\,cm$^{-3}$, a peak phase-space density of $\sim$$2 \times 10^{-2}$, and an elastic collision rate of about 3500\,s$^{-1}$. During the evaporation process the density stays roughly constant and the elastic collision rate decreases to $\sim$700\,s$^{-1}$ before condensation. The evaporation efficiency is very large as we gain at least 3 orders of magnitude in phase-space density for a loss of atoms by a factor of 10.


\begin{figure}
\includegraphics[width=85 mm]{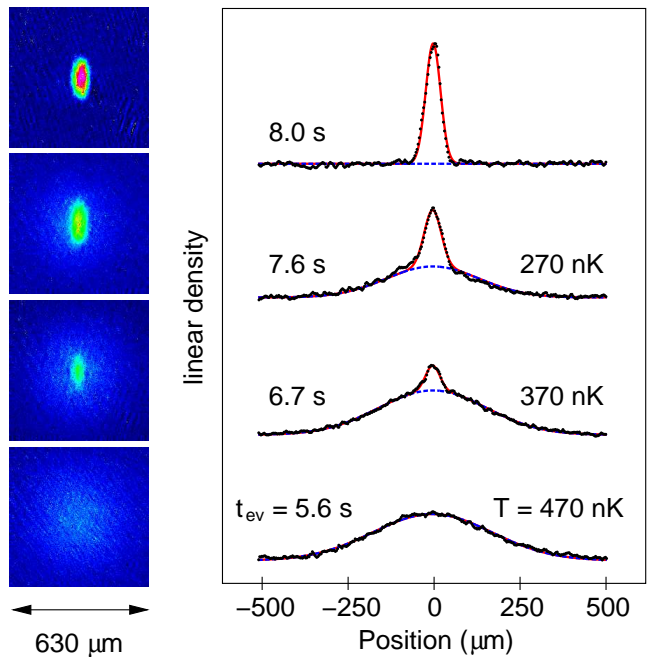}
\caption{\label{fig:Fig2Bimodal} Absorption images and integrated density profiles showing the BEC phase transition for different times $t_{\rm ev}$ of the evaporative cooling ramp. The images are along the vertical direction 25\,ms after release from the trap. The solid line represents a fit with a bimodal distribution, while the dashed line shows the Gaussian-shaped thermal part, from which the given temperature values are derived.}
\end{figure}

The phase transition from a thermal cloud to BEC becomes evident in the appearance of a textbooklike bimodal distribution, as clearly visible in time-of-flight absorption images and the corresponding linear density profiles shown in Fig.~\ref{fig:Fig2Bimodal}.
At higher temperatures the distribution is thermal, exhibiting a Gaussian shape. Cooling below the critical temperature $T_c$ leads to the appearance of an additional, narrower and denser, elliptically shaped component, representing the BEC. The phase transition occurs after 6.3\,s of forced evaporation, when the power of the horizontal beam is 190\,mW and the one of the vertical beam is 410\,mW. At this point, with the effect of gravitational sag taken into account, the trap depth is 2.8$\mu$K. The oscillation frequencies are 59\,Hz in the horizontal axial direction, 260\,Hz in the horizontal radial direction, and 245\,Hz in the vertical direction.

For the critical temperature we obtain $T_c = 420$\,nK by analyzing profiles as displayed in Fig.~\ref{fig:Fig2Bimodal}. This agrees within 20\%, i.e.\ well within the experimental uncertainties, with a calculation of $T_c$ based on the number of $3.8 \times 10^5$ atoms and the trap frequencies at the transition point. Further evaporation leads to an increase of the condensate fraction and we obtain a nearly pure BEC without discernable thermal fraction after a total ramp time of 8\,s.
The pure BEC that we can routinely produce in this way contains $1.5\times10^5$ atoms and its lifetime exceeds 10\,s.

\begin{figure}
\includegraphics[width=85 mm]{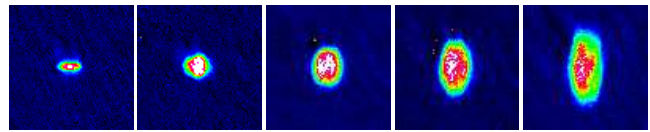}
\caption{\label{fig:Fig3Inversion} Inversion of the aspect ratio during the expansion of a pure BEC. The images (field of view 250$\mu$m $\times$ 250$\mu$m) are taken along the vertical direction. The first image is an in-situ image recorded at the time of release. The further images are taken 5\,ms, 10\,ms, 15\,ms, and 20\,ms after release.}
\end{figure}

The expansion of the pure condensate after release from the trap clearly shows another hallmark of BEC. Figure~\ref{fig:Fig3Inversion} demonstrates the well-known inversion of the aspect ratio \cite{Anderson1995oob, Inguscio1999book}, which results from the  hydrodynamic behavior of a BEC and the fact that the mean field energy is released predominantly in the more tightly confined directions. 
Our images show that the cloud changes from an initial prolate shape with an aspect ratio of at least 2.6 (limited by the resolution of the in-situ images) to an oblate shape with aspect ratio 0.5 after 20\,ms of free expansion. From the observed expansion we determine a chemical potential of $\mu/k_B\approx150$\,nK for the conditions of Fig.~\ref{fig:Fig3Inversion}, where the trap was somewhat recompressed to the setting at which the phase transition occurs in the evaporation ramp. Within the experimental uncertainties, this agrees with the calculated value of $\mu/k_B\approx180$\,nK.


It is interesting to compare our number of $1.5 \times 10^5$ atoms in the pure BEC with other BECs achieved in two-electron systems. For $^{174}$Yb up to $6\times 10^4$ atoms were reported \cite{Fukuhara2009aof}, and for $^{40}$Ca the number is $2\times 10^4$ \cite{Kraft2009bec}. It is amazing that our BEC clearly exceeds these values with little efforts to optimize the number after our first sighting of BEC (26 September 2009). We anticipate that there is much more room for improvement, in particular, in the transfer from the red MOT into the ODT. We interpret the amazing performance of $^{84}$Sr as the result of a lucky combination of favorable scattering properties with excellent conditions for narrowline cooling, and we believe that this makes $^{84}$Sr a prime candidate for future experiments on BEC with two-electron systems.

We finally discuss a few intriguing applications which seem to be realistic on a rather short time scale. The $^{84}$Sr BEC may serve as an efficient cooling agent to bring other isotopes into degeneracy. A BEC of $^{88}$Sr would be a noninteracting one \cite{Chin2008fri}, as the intraisotope scattering length is extremely small \cite{Ferrari2006cos, Martinezdeescobar2008tpp}. This would constitute a unique source of low-momentum, noninteracting, and magnetically insensitive atoms, ideal for precision measurements \cite{Sorrentino2006lca}. The fermionic isotope $^{87}$Sr offers a nuclear spin decoupled from the electronic degrees of freedom, which is very favorable for quantum computation \cite{Daley2008qcw, Gorshkov2009aem} and the essential key to a new class of many-body physics with ultracold atoms \cite{Hermele2009mio, Gorshkov2009tom}. The realization of a Mott insulator state appears to be a straightforward task \cite{Fukuhara2009mio}. Another fascinating application would be the creation of ultracold dimers made of an alkali-metal atom and a two-electron atom \cite{Nemitz2009poh}. Since all-optical evaporative cooling strategies for $^{87}$Rb and $^{84}$Sr proceed under very similar conditions \cite{Barrett2001aof, Cennini2003bec, Kinoshita2005aob}, the creation of SrRb molecules seems to be a realistic option. Such molecules would be qualitatively different from the bi-alkali-metal atoms currently applied in heteronuclear molecule experiments \cite{Krems2009book} as they offer a magnetic rovibrational ground state \cite{Nemitz2009poh}.

\begin{acknowledgments}
We thank Andreas Trenkwalder and Christoph Kohstall for technical assistance with the dipole trap laser. We furthermore thank Andrew Daley, Peter Zoller, and Hanns-Christoph N\"agerl for stimulating discussions on the prospects of strontium for future experiments. We also acknowledge fruitful discussions with Jun Ye, Thomas Killian, and Yoshiro Takahashi.
\end{acknowledgments}



\begin{thebibliography}{47}
\bibitem[{MOT()}]{MOTdetails}
\bibinfo{note}{Loading of the blue MOT follows standard laser cooling and
  trapping procedures. Atoms emitted from an effusive beam source are
  transversely laser cooled to increase the loading flux. They are then Zeeman
  slowed to be captured by the MOT. Our MOT laser beams have a waist of 9\,mm,
  a peak intensity of 5\,mW/cm$^2$, and a detuning of $-33$\,MHz. The
  quadrupole magnetic field has a gradient of 61\,G/cm along its symmetry axis.
  The temperature in the MOT is Doppler-limited to a few mK.}
\bibitem[{689()}]{689nmlaser}
\bibinfo{note}{The 689-nm laser system is a master-slave combination. The
  master laser is stabilized on a 100\,kHz linewidth cavity by optical
  feedback, resulting in a linewidth of $\sim$2\,kHz. The frequency is locked
  to the $^{88}$Sr intercombination line and shifted by 351.4\,MHz to the red
  to account for the isotope shift. A 30-mW slave laser diode is injection
  locked to the master. The red light is superimposed with the blue MOT beams
  using dichroic mirrors.}
\bibitem[{ver()}]{verticalnote}
\bibinfo{note}{For reasons of optical access the beam is not exactly vertical,
  but inclined by 20$^\circ$.}
\bibitem[{noi()}]{nointerference}
\bibinfo{note}{To avoid perturbations by interference effects the beams are
  orthogonally polarized and have a frequency difference of 160\,MHz.}
\bibitem[{rob()}]{robustramp}
\bibinfo{note}{The evaporation process turns out to be very robust, and common
  or individual variations of the ramping time constants for both beams are not
  found to have strong effects on the evaporation efficiency.}
\end{thebibliography}

\end{document}